\documentclass[proceedings]{JHEP3}

\PrHEP{PrHEP AHEP2003}

\usepackage{epsfig}  

\newcommand{\be}{\begin{eqnarray}}
\newcommand{\ee}{\end{eqnarray}}

\conference{International Workshop on Astroparticles and High Energy Physics}

\title{Loop level lepton flavor violation at linear colliders }

\author{\speaker{Mirco Cannoni}\\  
        Dipartimento di Fisica, Universit\`a degli Studi di Perugia and INFN Sezione di Perugia,\\   
        Via A. Pascoli 1, 06123, Perugia, Italy\\
        E-mail: \email{mirco.cannoni@pg.infn.it}}                       
\author{Orlando Panella and Stephan Kolb\\
        INFN, Sezione di Perugia, Via A.~Pascoli 1, 06123, Perugia, Italy}

\abstract{
We present a study of loop-level lepton flavor vioalting signals in 
models with heavy Majorana neutrinos and in the supersymmetric extension of the standard model. 
The attention is focused to the $e^- e^-$ option of the next generation of linear colliders and 
its potential of discovering new physics is emphasized.}

\begin{document}

\section{Introduction}

Linear $e^+ e^-$ colliders are very important step towards the understanding of high energy particle
interactions.
Besides confirming and allowing precision measurements on new physics which, hopefully, will be discovered at the LHC,
they offer by themselves new physics opportunities thanks to the possibility of
$\gamma \gamma$, $\gamma e^{-} $, $e^- e^-$ collisions. These options will allow to test, with 
higher sensitivity couplings of the standard model (and alternative theories) which cannot be studied in hadronic or $e^+ e^-$
collisions. Moreover the possibility of employing electron beams with a high degree of longitudinal 
polarization will be essential to enhance signals of new physics. 

The $e^- e^-$ mode with lepton number $L=+2$ of the initial state is particularly suitable 
for studying lepton and flavor number violating reactions (LFV). 
The ``merits'' of this option have been recently recalled by the ECFA/DESY working group in Ref.~\cite{DeRoeck}.
Among these there is (1) the possibility to identify Majorana neutrinos of masses in the TeV range through the ``inverse
neutrinoless double beta decay'' reaction $e^- e^- \to W^- W^-$, and  
(2) the sharper onset of slepton production threshold respect to the $e^+ e^-$ option.
Here we present a study of two loop-level reactions strictly connected with these two ``merits'':
we first consider seesaw type models with heavy Majorana neutrinos at the TeV scale
and study the reactions $e^{-}e^{-} \to \ell^{-}\ell^{-}$,
and then similar reactions $e^{-}e^{-} \to \ell^{-}e^{-}$ ($\ell=\mu,\tau$) in supersymmetric
models where LFV is due to slepton mixing. The importance of the $e^- e^-$ mode for studies 
of LFV due to slepton mixing and the advantages of the threshold behavior was first noted by Feng in various 
papers~\cite{Feng1,Feng2,Feng3,Feng4,Feng5}.  

We briefly discuss the standard model background. More details on the calculations and numerical
tools used can be found in Refs.~\cite{Cannoni2002,Cannoni2003}.  

\section{$e^{-}e^{-}\;\to\;\ell^{-}\ell^{-}$ ($\ell=\mu,\tau$) through heavy Majorana neutrinos}

To give a detectable signal, heavy Majorana neutrinos (HMN), besides having masses in the TeV range, 
must have
interactions which are not suppressed by the mixing matrices with light states as instead happens in the one 
family seesaw mechanism, where $\theta\ {\simeq}\ \sqrt{{m_{\nu}}/{M_N}}$. 
With three generations, more free parameters are at our disposal, and the ``two miracles'' of not so large masses 
and non negligible mixing, are obtained imposing suitable relations among the elements of the matrices $m_D$
and $M_R$: examples of these models were proposed some time ago in Refs.~\cite{buchmuller91,chang,heusch}
and in the more recent paper,
Ref.~\cite{loinaz}, whose authors suggest ``neutrissimos'' as the correct name for these particles.
According to other authors, Ref.~\cite{gluza2000}, these models are based on fine-tuned relations, but are shown 
to be not 
in contradiction with any experimental bound. We do not enter in such a theoretical dispute and assume
``neutrissimos'' in the TeV range and study the phenomenological consequences.

Experimentally one cannot put bounds on the single mixing matrix elements, but on some combinations 
of them, assuming that each charged lepton couples only to one heavy neutrino with significant strength.
Light-heavy mixing has to be inferred
from low-energy phenomenology
and from global fits performed
on LEP data identifying the following effective mixing angles 
$s^{2}_{\ell} =\sum_{j} |B_{\ell_i N_j}|^{2}{\equiv}\sin^{2}{\theta}_{{\nu}_{\ell}}$
with upper bounds~\cite{Nardi}:
\begin{eqnarray}
s^{2}_{{e}}< 0.0054,\;\;\;
s^{2}_{{\mu}}< 0.005,\;\;\;
s^{2}_{{\tau}}< 0.016,
\label{mixing}
\end{eqnarray}
where $B$ is the mixing matrix appearing in the charged current weak interaction lagrangian.
Under these assumptions, the coupling of neutrissimos to gauge bosons and leptons is numerically fixed 
to $gB_{\ell N_{i}}$,
where $g$ is the $SU(2)$ gauge coupling of the SM. 
Since the width of the heavy states grows as $M^{3}_N$, at a certain value it will 
happen that $\Gamma_N > M_N$, signaling a 
breakdown of 
perturbation theory. 
The perturbative limit on $M_N$ is thereby estimated requiring 
$\Gamma_N < M_N/2$, which gives an upper bound of 
$\simeq 3$ TeV~\cite{Illana} for the numerical values given in Eq.~(\ref{mixing}).

We study the process $e^{-}e^{-}\to \ell^{-}\ell^{-}$ ($\ell=\mu,\tau$),
which violates the $L_e$ and $L_{\ell}$ lepton  numbers.
\DOUBLEFIGURE[t]
{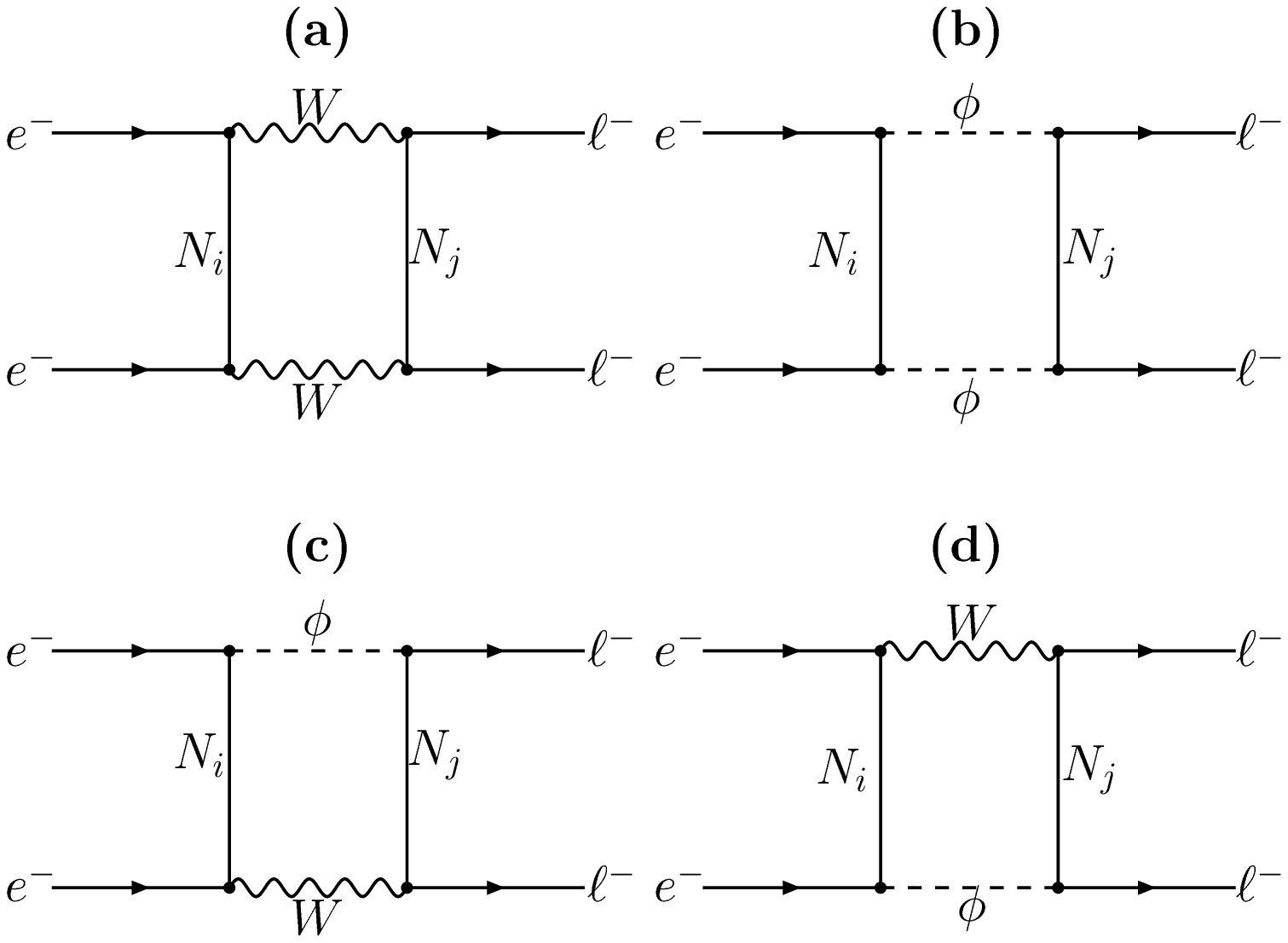,width=7.3cm,height=8cm,clip=}
{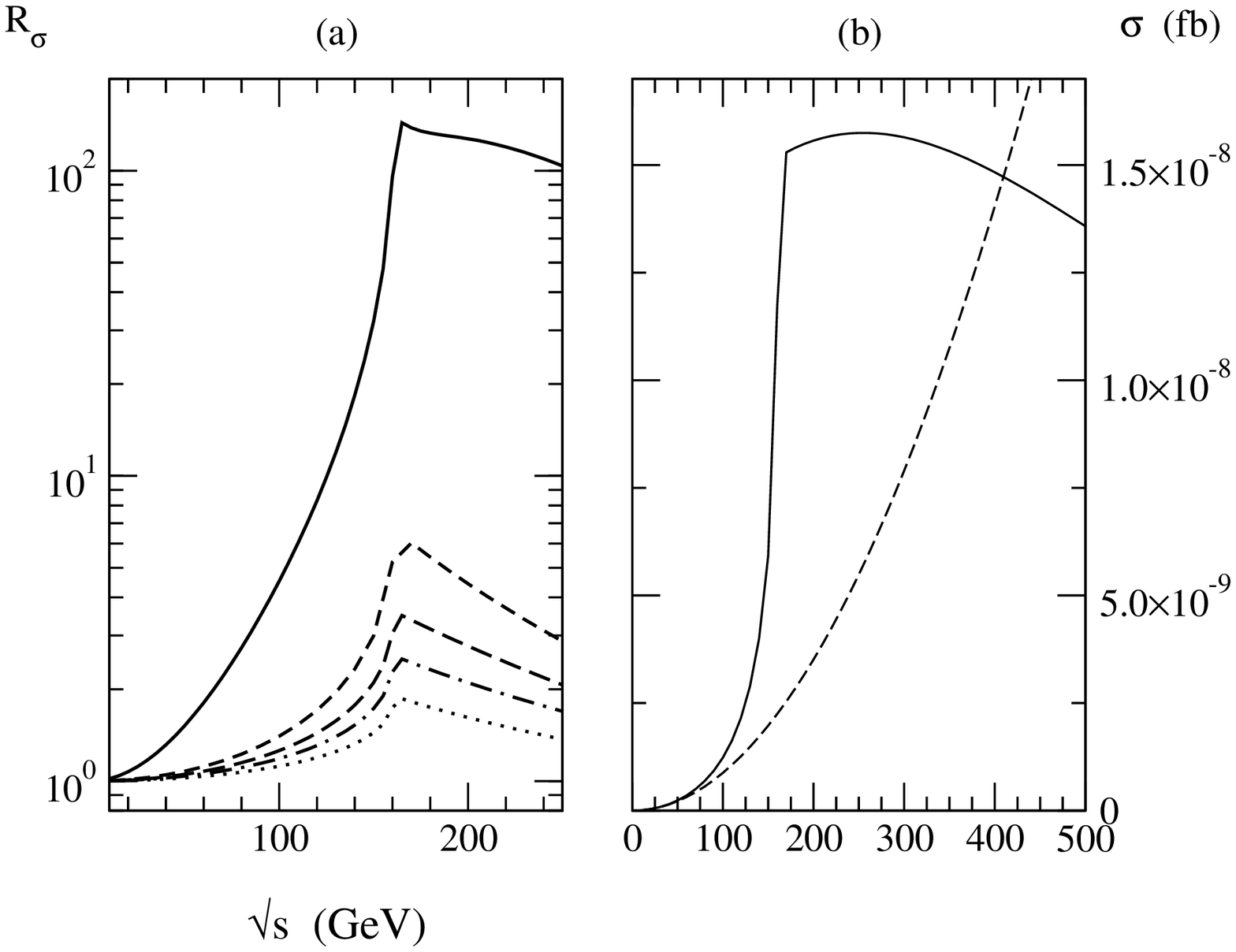,width=7.3cm,height=8cm,clip=}
{In (a-d) the Feynman diagrams, in the 't Hooft-Feynman 
gauge, contributing to $e^-e^- \to \ell^-\ell^- $ ($\ell=\mu,\tau$) via heavy Majorana neutrinos.
$\phi$ is the unphysical Goldstone boson.~\label{FIG1}} 
{{\bf(a)}the ratio $R_\sigma$ is plotted as a function of $\sqrt{s}$. {\it Solid line}, 
$M_{N_i}=M_{N_j}=100$ GeV; 
{\it dotted}, $M_{N_i}=M_{N_j}=3$ TeV.
{\bf(b)}{\it Solid line}, our calculation, {\it dashed line} that of Ref.~\cite{Pham:2000bz}.~\label{FIG2}}
The amplitudes are computed in the 't Hooft-Feynman gauge where
there are graphs with $WW$, $\phi\phi$ and
$\phi{W}$ exchange, $\phi$ being the Goldstone boson, as shown in Fig.~\ref{FIG1}.
The numerical computation of the four-point functions was performed 
using the 
{\scshape{looptools}}~\cite{looptools} package. 
A similar study was done in Ref.~\cite{Pham:2000bz} using the approximation where {\em all external 
momenta in the loops are  neglected} relative to the heavy masses 
of the gauge bosons and Majorana neutrinos, that is $q_i = k+k_i\simeq k$,  
where $k$ is the integration variable in the loop integrals and $k_i$ are the sums
of external momenta appearing in the propagators.
This approximation for the four-point functions is good at low energies, 
such as in decay processes of heavy mesons,
or when $\sqrt{s}<<M$, $M$ being the highest mass running in the loop. 
In this way the cross section presents a linear  
growth with $s$ which breaks unitarity: therefore, in order
to make quantitative predictions with the correct high energy behavior, 
full dependence on the external momenta of the four-point functions has to be considered. 
Theoretically, according to the `Cutkosky rule', 
one expects an enhancement at $\sqrt{s}\simeq 161$ GeV $\simeq{2M_{W}}$,
the threshold for on-shell $W W$ gauge boson production, at which 
the four-point functions develop an imaginary part.
\DOUBLEFIGURE
{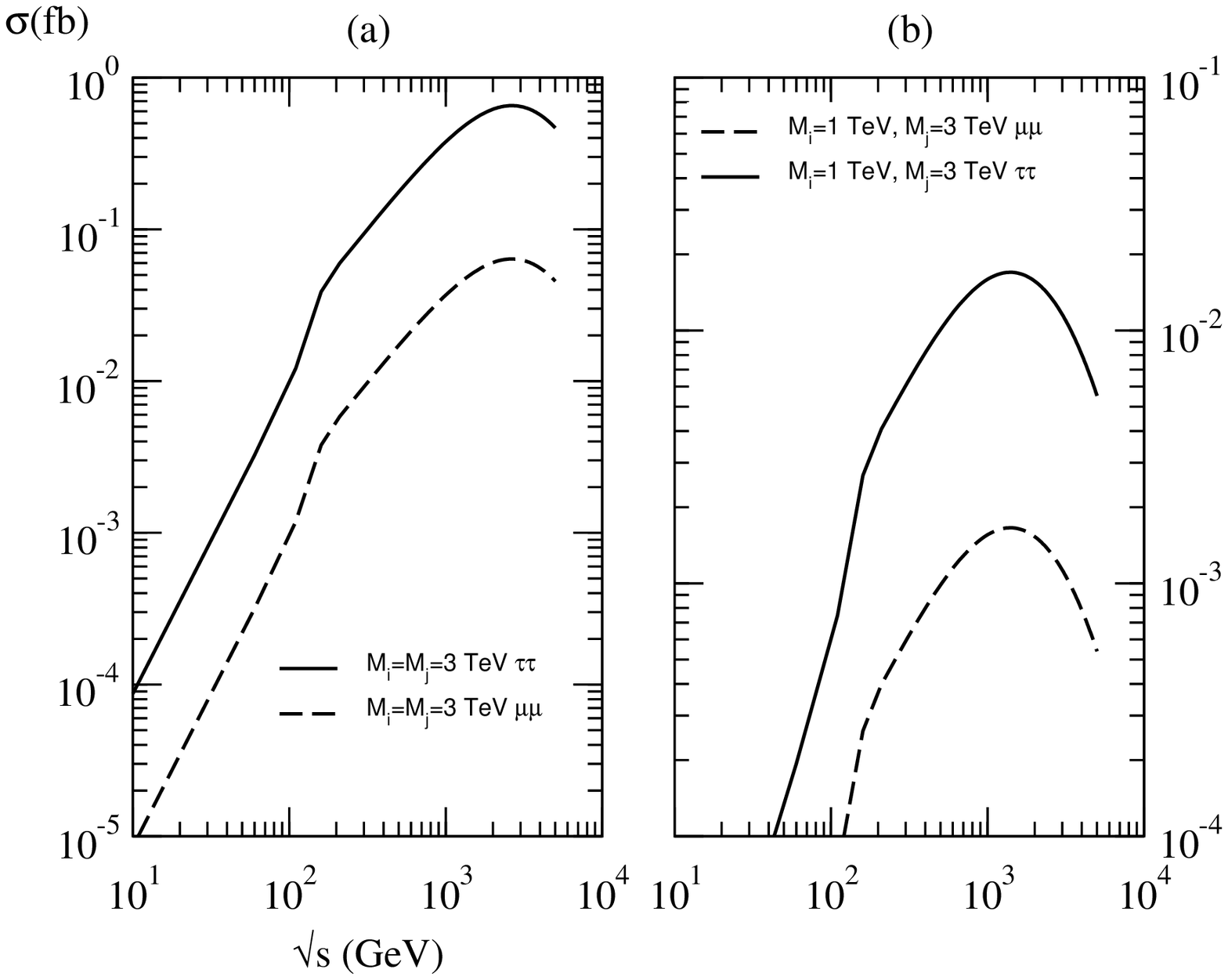,width=7.3cm,height=8.5cm,clip=}
{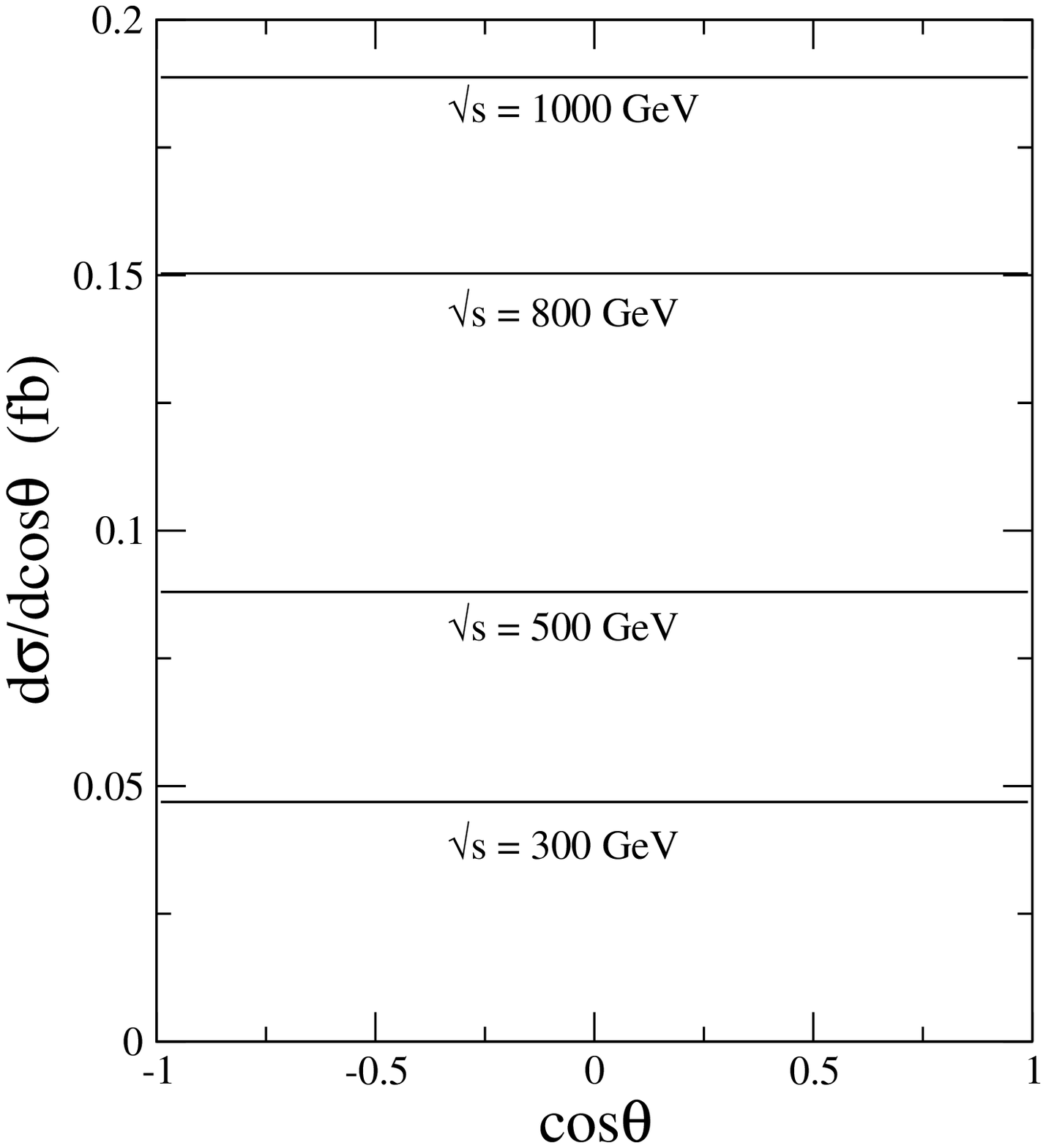,width=7.3cm,height=8.5cm,clip=}
{Total cross sections as function of $\sqrt{s}$. 
{\bf(a)} the solid curve referes to the case of 
$e^-e^- \to \tau^-\tau^-$ with $M_{N_i} = M_{N_j}= 3$ TeV,
while the dashed line referes to  ($ e^-e^- \to \mu\mu$) with the 
same values of Majorana masses.
{\bf(b)} the Majorana masses are changed to 
somewhat lower values: $M_{N_i} = 1$ TeV, $ M_{N_j}= 3$ TeV.\label{FIG3}}
{Angular distribution in the polar angle of the outgoing lepton for
different values of the center of mass energy, $\sqrt{s}$ 
in the case of $e^- e^- \to \tau^- \tau^-$ with $M_{N_i} = M_{N_j}= 3$ TeV.
The curves are not exactly constant, and using an appropriate  scale 
they show small deviations from a stright line, remaining left-right symmetric.\label{FIG4}}
In Fig.~\ref{FIG2}(a) the ratio $R_\sigma = \sigma_{tot}/ \sigma_0$
of the integrated total cross section 
$\sigma_{tot}$ to $\sigma_0$, the cross section of the low 
energy calculation of Ref.~\cite{Pham:2000bz}, is plotted for sample values of 
the Majorana masses. 
The enhancement due to the threshold singularity of the loop amplitude
is more pronounced for values of Majorana masses close to $M_W$ and 
is drastically reduced increasing $M_{N_i} \approx M_{N_j}$ to  
${\cal O}$(TeV). 
As $R_\sigma \to 1$ at $\sqrt{s} <<  M_W$ in all the cases, 
the agreement of our full calculation 
with the result of Ref.~\cite{Pham:2000bz} in the regime of 
low energies is evident.
The threshold effect appears to be quite spectacular 
only for values of Majorana 
masses which correspond to cross sections too small to be measured 
even at a next linear collider. 
In Fig.~\ref{FIG2}(b)
the effect of the threshold singularity in the loop integral is shown 
reporting absolute cross sections for a particular choice of Majorana masses: 
$M_{N_i} = 150$ GeV and $M_{N_j} = 450$ GeV. 
The low energy 
approximation (dashed line), obtained neglecting external momenta 
in the loop, is inadequate when 
the energy of the reaction increases to values comparable with the masses.
Increasing the energy, after reaching a maximum, 
the cross section starts to decrease until the asymptotic behaviour 
${\cal O}(1/{s^2})$ of the loop integral is reached. 
This happens for every value of 
heavy Majorana neutrino masses and we checked numerically that, as expected,
for higher masses the asymptotic regime is reached at 
higher values of $\sqrt{s}$. 
In fact from Fig.~\ref{FIG3} we note that the cross section grows with increasing 
HMN masses. 
The main contribution comes 
from the graph with two Goldstone bosons since  
their coupling is proportional to $M_{{N}_{i}}$. Moreover the chiral 
structure of the coupling selects the 
mass term in the numerator of the Majorana neutrino propagators. 
When these masses are much larger then the other quantities, 
the amplitude scales like $M^{3}_{{N}_{i}} 
M^{3}_{{N}_{j}}/M^{2}_{{N}_{i}}M^{2}_{{N}_{j}}\simeq M_{{N}_{i}}M_{{N}_{j}}$, 
i.e. is proportional to the square of the heavy masses.
This fact is  the well known non decoupling of heavy fermions 
in theories with spontaneous symmetry breaking (similarly in the SM 
the top quark gives sizable radiative corrections owing to its large 
mass and  a quadratic non decoupling).
In Fig.~\ref{FIG3}(a) the cross section is plotted for masses up to 
the perturbative limit, using the maximally allowed value of the mixing. 
We see that for $M_{N_i} = M_{N_j} = 3 $ TeV the signal does reach the 
level of $10^{-1},10^{-2}$ fb respectively 
for the ($\tau\tau$) and the ($\mu\mu$) signals at $\sqrt{s}= 500 $ GeV, 
which for an annual integrated luminosity of $100$ fb$^{-1}$ would 
correspond respectively to 10 and 1 event/year. At higher energies,
$\cal O$ (TeV), one could get even larger event rates (30 and 3) respectively. 
The solid curve refers to $e^-e^- \to \tau^- \tau^-$: this is largest 
because the upper limits on the mixing are 
less stringent. One can also see the onset of the asymptotic regime at 
$\sqrt{s}\approx 3 $ TeV.          
Fig.~\ref{FIG3}(b) shows that the cross section quickly decreases as lower
Majorana masses are considered. 

As even in the more optimistic cases event rates are quite modest  
it is important to check how the signal cross-section is affected by 
kinematic cuts on the angle of the outgoing leptons.  
The 
angular distributions turn out to be 
practically constant as shown in Fig.~\ref{FIG4}. 
They are forward-backward symmetric because both 
the $t$ and $u$ channel are present. The absence of a strong 
dependence on the polar angle is due to the fact that within the range 
of the parameters  used here the contributing four point functions 
depend very mildly on the kinematic variables ($u$ and $t$). 
This behaviour can be most easily  understood using helicity 
amplitudes. 
In the limit of massless external particles the process is dominated by a well defined 
helicity amplitude: $ e_L e_L \to \ell_L \ell_L$.
In the center of mass frame this is a S-wave scattering with $J_z=0$,
meaning that the scattered particles are emitted back to back but 
without a preferred direction relative to the collision axis ($z$).   
Thus this signal is characterized by practically flat angular distributions 
and as a result the total cross section is quite insensitive 
to angular cuts.
We have verified that with
$|\cos{\theta}|{\le}0.99$ the change  in
$\sigma_T$ is $\approx 1\% $ for all energies considered, while using 
$|\cos{\theta}|{\le}0.95$  the total cross-section decreases by 
$\approx 5\%$. The reduction of the total cross section is 
measured almost precisely  by the reduction of the phase space, 
meaning that the angular distribution is constant up to $\approx 0.1\%$. 
Thus it can be concluded that 
the number of events will not be drastically affected for any 
reasonable choice of experimental cuts.  
The possibility of employing electron beams with a high degree of longitudinal 
polarization as planned for the new colliders will be essential to discover rare signals: in our case 
only an helicity amplitude contribute, so that using left-polarized electron beams will single out the
essential spin configuration and we can gain a factor $\simeq 4$ on the previous cross sections,
because the avarage on the initial spins is not needed. 
This consideration applies also in the next Section.

\section{$e^{-}e^{-}\;\to\;\ell^{-}e^{-}$ ($\ell=\mu,\tau$) in R-conserving supersymmetric models}

Slepton and squark mass matrices in soft-breaking potential, for example $m^2_{\tilde{L}}$, 
are in general complex non-diagonal matrices. The diagonalization implies the presence of generational  mixing matrices 
at the lepton-slepton-gaugino vertices.
These couplings can originate too high rates for rare unobserved flavor changing processes. The SUGRA boundary conditions
and the renormalization group equations (RGE) evolution originate flavor-diagonal sfermion mass matrices, 
so besides offering a mechanism for reducing
to a manageable number the soft parameters, also cure the 'flavor problem'. On the other hand, if one wants to study 
the phenomenology of the MSSM without referring to a particular high energy 
scenario, one can consider general mass matrices      
and accept their flavor violating entries as large as the experimental bounds allow them. 

However when the seesaw mechanism 
is embedded in the MSSM with mSUGRA boundary conditions at high energy and heavy Majorana neutrinos 
are out of the colliders reach, a new source of LFV arises.
The seesaw mechanism requires that the superpotential
contains three $SU(2)_L$ singlet neutrino superfields $\hat{N_{i}}$ with the following 
couplings~\cite{borma,hisa2}:
\begin{eqnarray}
W=(Y_{\nu})_{kl}\varepsilon_{ij}\hat{H}_{2}^{i}\hat{N}_k \hat{L}^{j}_l
+\frac{1}{2}(M_{R})_{ij} \hat{N}_i \hat{N}_j.
\label{yuk}
\end{eqnarray}
Here $H_2$ is a Higgs doublet superfield, $L_i$ are the $SU(2)_L$ doublet lepton superfields, 
$Y_{\nu}$ is a Yukawa coupling matrix and $M_R$ is the $SU(2)_L$ singlet neutrino mass matrix.
With the additional Yukawa couplings in Eq.~(\ref{yuk}) and a new mass scale ($M_R$) 
the RGE evolution of the parameters is modified. 
Assuming that $M_{R}$ is the mass scale 
of heavy right-handed neutrinos, the RGE evolution from the GUT scale down to $M_R$ induce
off-diagonal matrix elements in $(m^{2}_{\tilde{L}})_{ij}$. 
At the GUT scale we assume the universal conditions 
$m^{2}_{\tilde{L}}=m^{2}_{\tilde{\nu}}=m^{2}_{H_2}=m^{2}_{0}$ and $A_{\nu}=a m_{0}Y_{\nu}$,
thus the correction of the off-diagonal elements are (in the leading-log approximation):
\begin{eqnarray}
(m^{2}_{\tilde{L}})_{ij}\simeq -\frac{1}{8\pi^{2}}(3+a^{2}_{0})m_{0}^{2}
(Y_{\nu}^{\dagger}Y_{\nu})_{ij}\ln\left(\frac{M_{GUT}}{M_{R}}\right),
\label{dl2}
\end{eqnarray} 
which depends crucially on the non-diagonal elements of $(Y_{\nu}^{\dagger}Y_{\nu})_{ij}$, the square of
the neutrino Yukawa couplings. 
The main point is that these elements can be large numbers because in 
the seesaw mechanism they do not directly determine the mass of the light neutrino, but only through the
seesaw relation $m_\nu \simeq m^2_D / M_R =v^2 Y^2_\nu/ M_R$. 
On the other hand the same effect on the mass matrix of $SU(2)_L$ singlet charged 
sleptons $(m^{2}_{\tilde{R}})_{ij}$ is 
smaller: in fact  
in the same leading-log approximation of Eq.~(\ref{dl2}), the corresponding RGE do not contain terms 
proportional to $Y_{\nu}^{\dagger}Y_{\nu}$, since the right-handed leptons fields have only 
the Yukawa coupling $Y_{\ell}$, which completely determines the Dirac mass of the charged leptons and these 
are known to be small numbers.
Thus the off-diagonal elements $(m^{2}_{\tilde{R}})_{ij}$ can be taken
to be $\simeq 0$. 
The slepton mass eigenstates are obtained diagonalizing the slepton 
mass matrices. 
The corresponding mixing matrices induce LFV couplings in the 
lepton-slepton-gaugino vertices  
$\tilde{\ell}^{\dagger}_{L_{i}}{U_{L}}_{ij}{\ell}_{L_{j}}\chi$. 
The magnitude of LFV effects will depend on the RGE induced non diagonal entries and 
ultimately on the neutrino Yukawa couplings $(Y_{\nu})_{ij}$. 

The rate of LFV transitions like $\ell_{i} \to \ell_{j}$, $i\neq j$, 
$\ell=e,\mu,\tau$ induced by the lepton-slepton-gaugino vertex is determined by the
mixing matrix ${U_{L}}_{ij}$ which, as stated above, is model dependent.
In a model independent way, however, one can take the 
lepton, slepton, gaugino vertex flavor conserving with the slepton 
in gauge eigenstates, so that LFV is 
given by mass insertion of non diagonal slepton propagators. 
In a similar spirit our phenomenological approach
will be quite model independent and, in order to keep the discussion simple enough, 
the mixing of only two generations will be considered.
Thus the slepton and sneutrino mass matrix are:
\begin{eqnarray}
\tilde{m}^{2}_{{L}}=\left(\begin{array}{cc}
 \tilde{m}^{2} & \Delta m^2\\
       \Delta m^2 & \tilde{m}^{2}
\end{array}\right),
\end{eqnarray}
with eigenvalues: $\tilde{m}^{2}_{\pm}=\tilde{m}^2\pm \Delta m^2$ 
and maximal mixing matrix.  
Under these assumptions the LFV propagator in momentum space for a scalar line is
\begin{eqnarray}
\langle\tilde{\ell}_{i}\tilde{\ell}^{\dagger}_{j}\rangle_{0}=
\frac{i}{2}\left(
\frac{1}{p^{2}-\tilde{m}^{2}_{+}}-\frac{1}{p^{2}-\tilde{m}^{2}_{-}}\right)
=i
\frac{\Delta m^2}{(p^{2}-\tilde{m}^{2}_{+})(p^{2}-\tilde{m}^{2}_{-})},
\label{LFVprop} 
\end{eqnarray} 
and the essential parameter which controls the LFV signal is
\be 
\delta_{LL}=\frac{\Delta m^{2}}{\tilde{m}^{2}}.
\label{lfvpar}
\ee
\DOUBLEFIGURE
{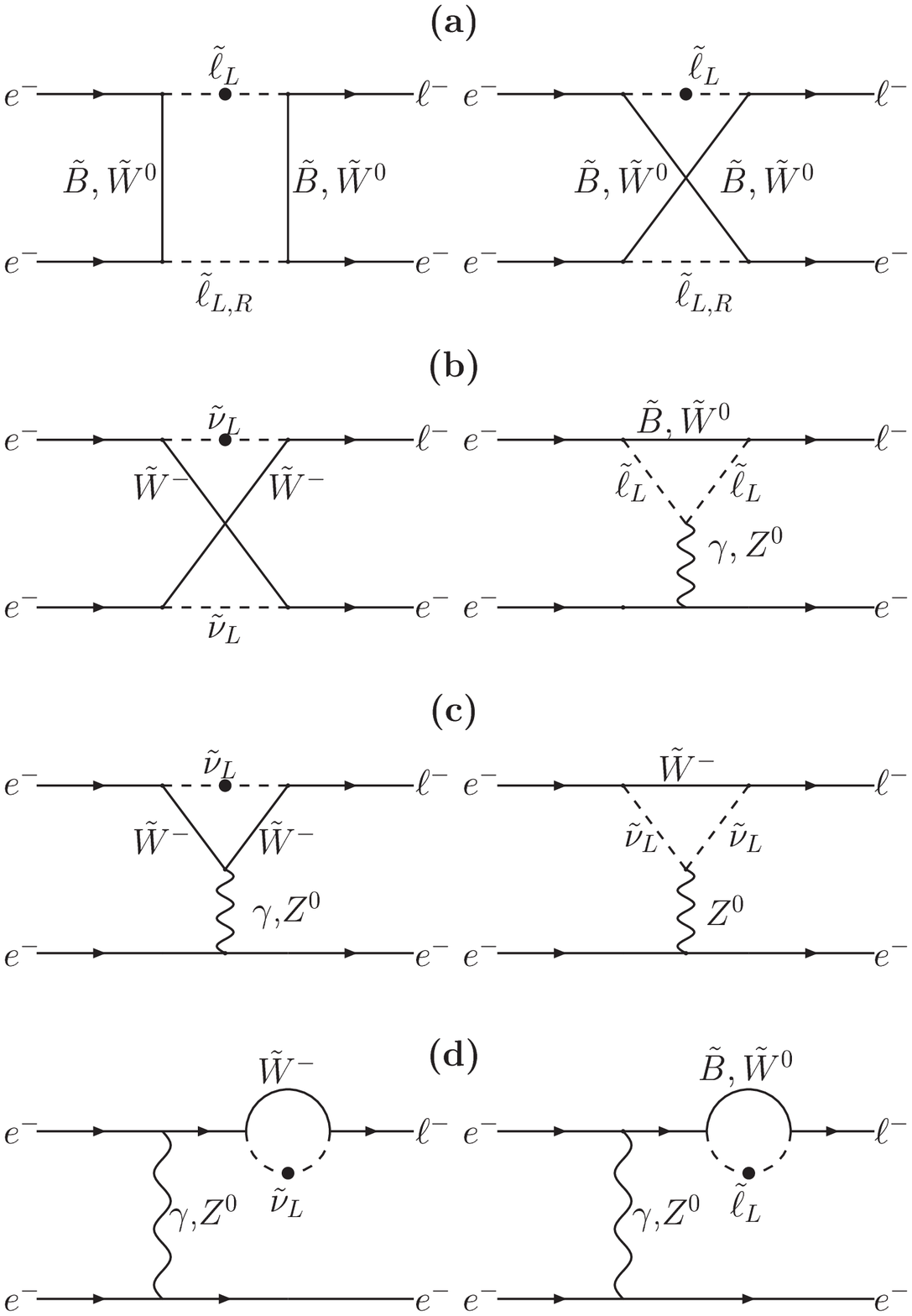,width=7.3cm,height=8.5cm,clip=}
{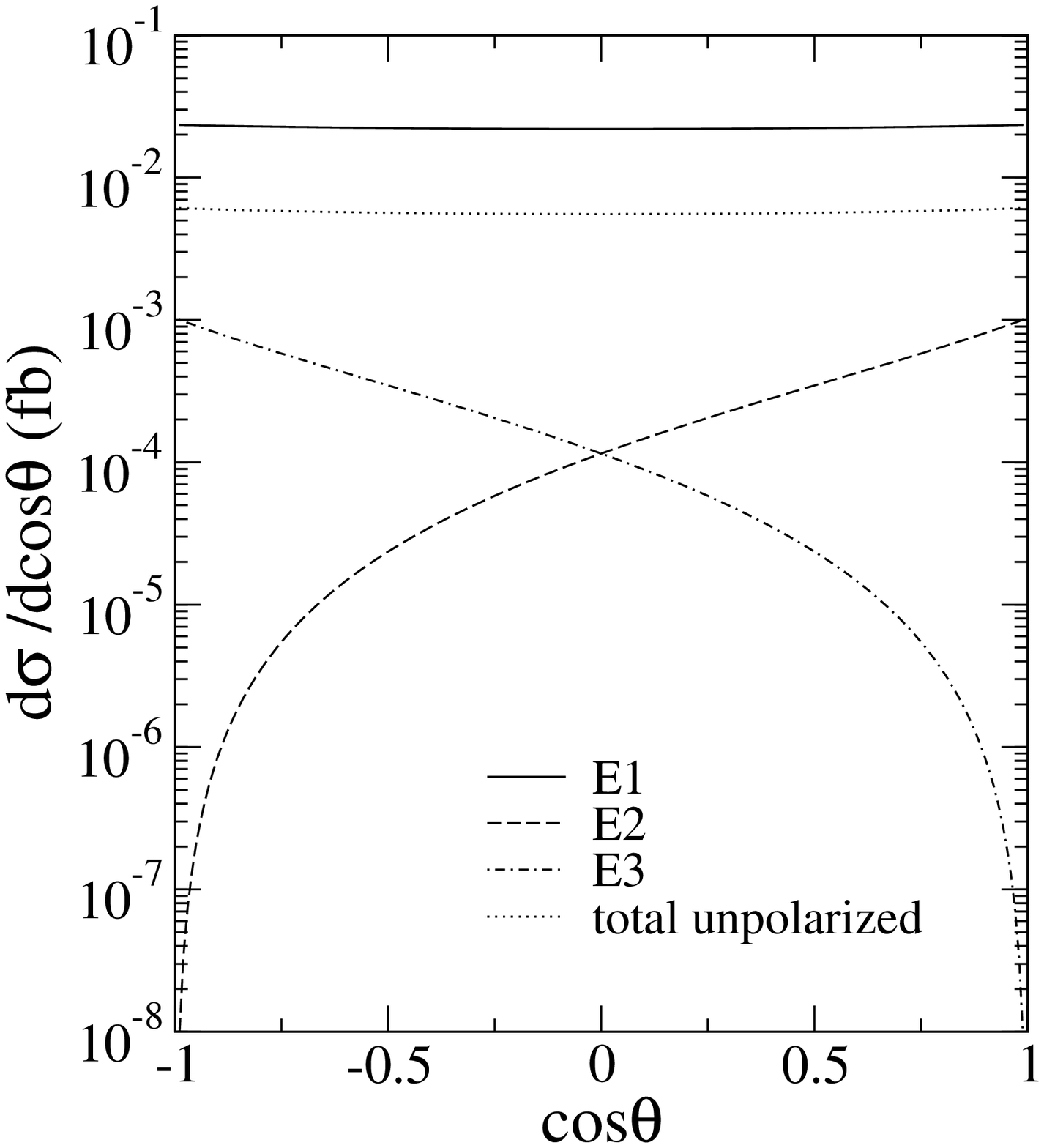,width=7.3cm,height=8.5cm,clip=}
{Feynman diagrams for $e^{-}e^{-}$ collisions. The full black dot 
in a scalar line denotes again the lepton flavour violating propagator. 
Exchange diagrams are not shown.\label{figee}}
{Differential cross sections as a function of the scattering angle. 
Values of the parameters:
$M_1=80,\; M_2=160,\;m_{\tilde{\ell}}=m_{\tilde{\nu}}=100$ GeV and 
$\Delta m^{2}=6000$ $\mbox{ GeV}^2$.   \label{eediff}}
In this scenario we consider the reaction
\be
e^-(p_1,\lambda_1)e^{-}(p_{2},\lambda_2) \to
\ell^{-}(p_3,\lambda_3)e^-(p_4,\lambda_4).
\ee 
whose Feynman diagrams are shown in Fig.~\ref{figee}.
Here $\lambda_i$ denotes the helicity of particle $i$. 
The total unpolarized cross-section (averaged over initial spins) is $\sigma=(1/4)\sum_j \sigma_j$.
The signal is suppressed if neutralinos and charginos $\chi^{0,\pm}$ 
are Higgsino-like, since their coupling is 
proportional to the lepton masses. 
For the same reason left-right mixing in the slepton matrix is neglected.
Therefore it is assumed that
the two lightest neutralinos are pure Bino and pure Wino 
with masses $M_1$ and $M_2$ respectively, while 
charginos are pure charged Winos with mass $M_2$, $M_1$ and $M_2$  
being the gaugino masses in the soft breaking potential.
Numerical results are obtained using the mSUGRA relation 
$M_1 \simeq 0.5 M_2$ for gaugino masses while $\Delta m^{2}$ and the slepton 
masses are taken to be free phenomenological parameters. 
The parameter space is scanned in order to identify the regions which may 
deliver an interesting signal. 
The contributing amplitudes are
\be
{\cal M}_{E1}&=&{\cal M}(e^-_L e^-_L \to \ell^-_L e^-_L),\cr
{\cal M}_{E2}&=&{\cal M}(e^-_L e^-_R \to \ell^-_L e^-_R),\cr
{\cal M}_{E3}&=&{\cal M}(e^-_R e^-_L \to \ell^-_L e^-_R).
\ee
The corresponding differential cross sections are plotted in 
Fig.~\ref{eediff}. 
${\cal M}_{E1}$ has $J_z=0$, is flat and forward-backward 
symmetric because of the antisymmetrization.
${\cal M}_{E2}$ and ${\cal M}_{E3}$ describe
P-wave scattering with $J_z=+1$ and $J_z=-1$ respectively: in order to 
conserve angular momentum ${\cal M}_{E2}$ must be peaked in the 
forward direction while ${\cal M}_{E3}$ favours backward scattering.
Both ${\cal M}_{E2}$ and ${\cal M}_{E3}$ 
are orders of magnitude smaller than ${\cal M}_{E1}$. 
The signal cross section is to a very good approximation given 
by the amplitude ${\cal M}_{E1}$. Since it is almost flat
the angular integration will give a factor almost exactly equal to two. 
This again shows  the importance of the option of having polarized beams.
If both colliding electrons are left-handed one singles out 
the dominant helicity amplitude and a factor four is gained
in the cross section relative to the unpolarized case.
This may be important in view of the relatively small signal 
cross section one is dealing with.
\DOUBLEFIGURE[t]
{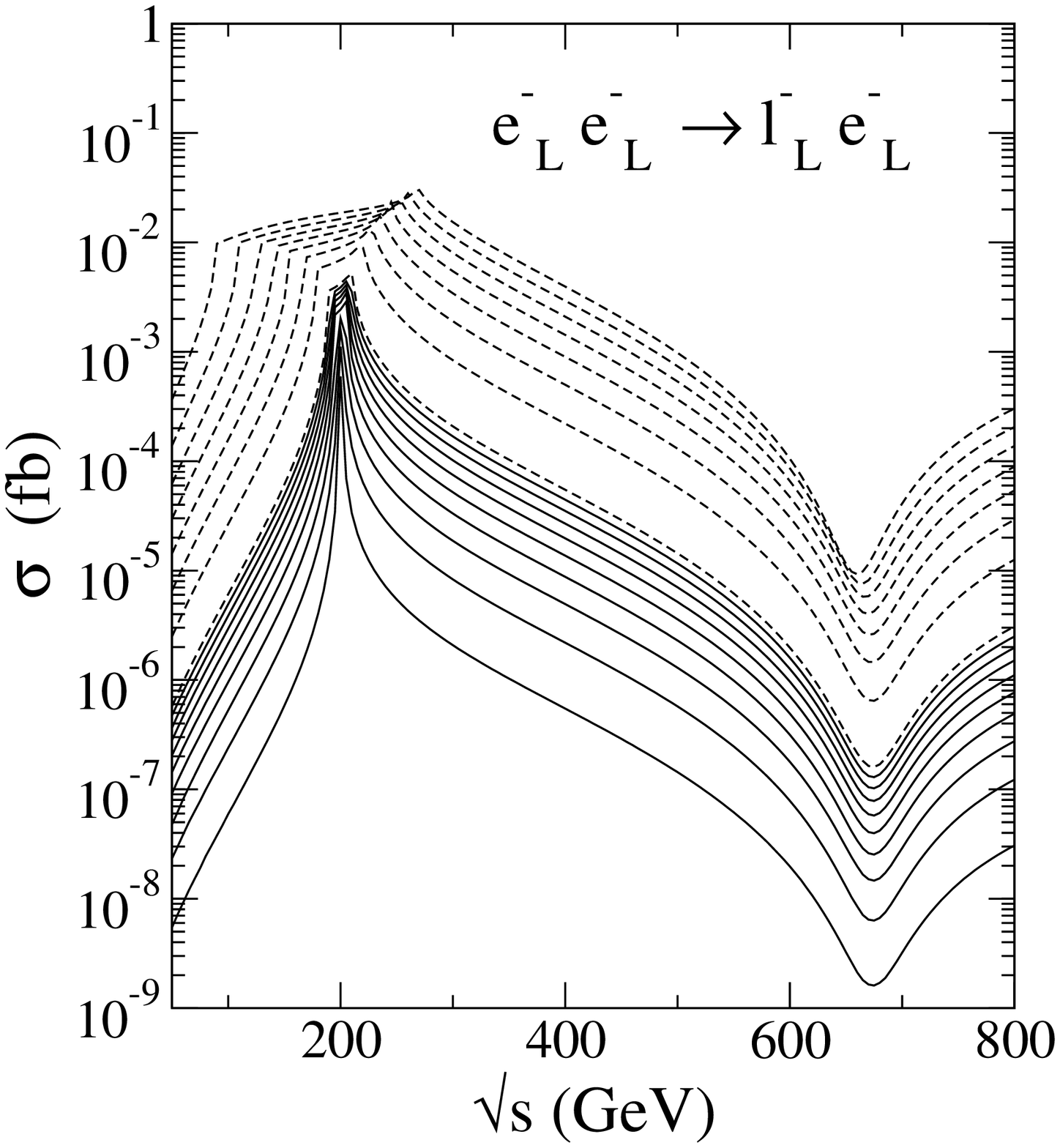, width=7.3cm,height=8.5cm,clip=}
{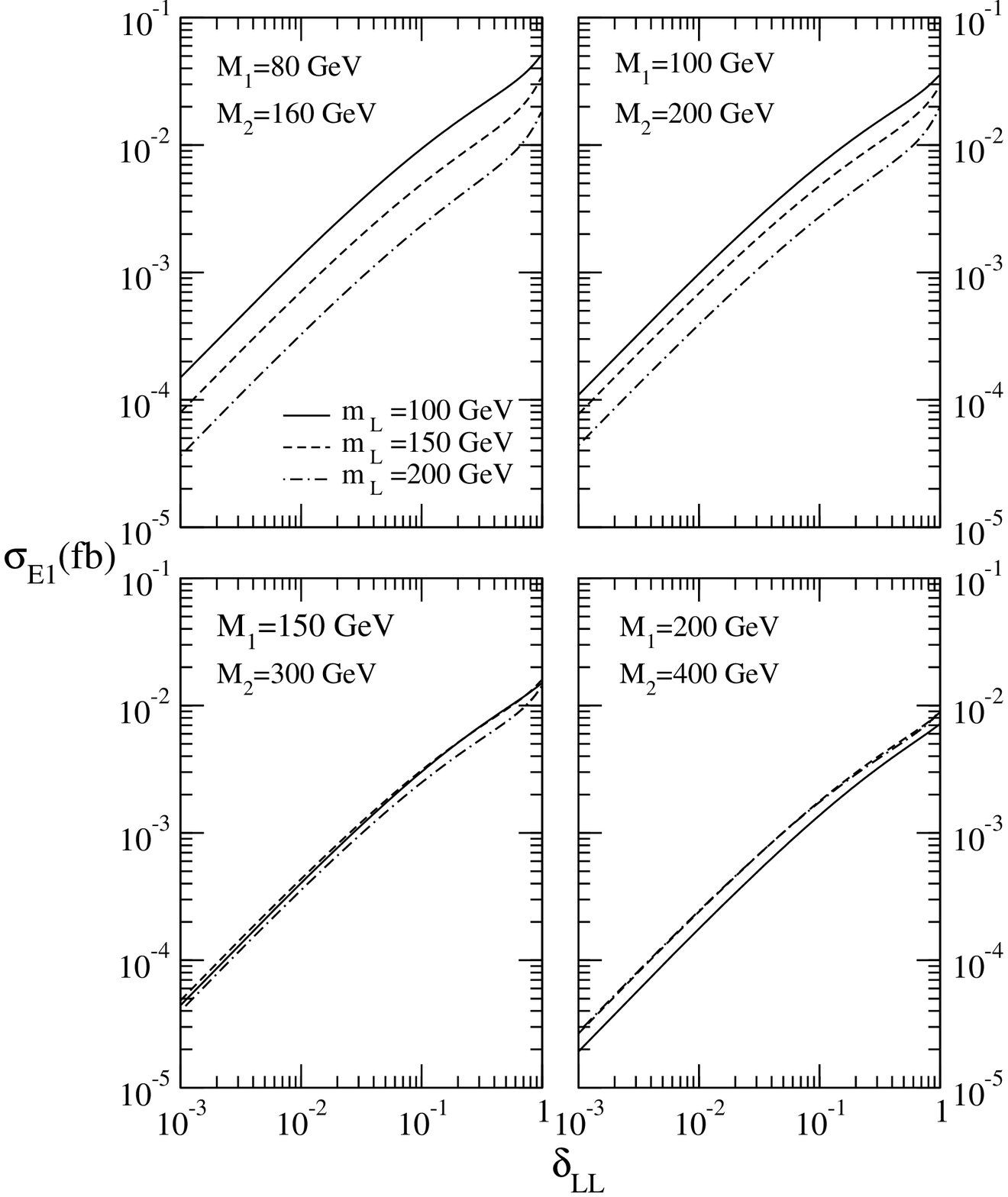, width=7.3cm,height=8.5cm,clip=}
{$\sigma_E1$ as a function of $\sqrt{s}$ with slepton and gaugino masses as in Fig.~6.
{\it Solid lines}: $\Delta m^2$ increasing from $100$ GeV$^{2}$
to $900$ GeV$^{2}$ in steps of $100$. {\it Dashed lines}:
from $1000$ to $8000$ GeV$^{2}$ in steps of $1000$.\label{eetot}}
{Total cross section for the amplitude E1 in function of 
$\delta_{LL}$ for $\sqrt{s} =2 {\tilde m}_L$.
The values of the other parameters are given in the legends.\label{eedelta}} 
The analysis of the total 
cross section as a function of $\sqrt{s}$ is the following (see Fig.~\ref{eetot}): 
the box diagrams dominate
at $\sqrt{s}=2\tilde{m}_L$ where $\sigma$ changes of orders of magnitude 
giving a sharp peak that is smeared only by large values of 
$\Delta m^2$, while penguin diagrams give a substantial contribution only at 
higher energies.
This can be easily understood considering the threshold behavior of the 
cross section for slepton pair production~\cite{Feng3,Peskin}: 
defining $\beta=\sqrt{1-4 m^{2}_{\tilde{L}}/s}$ the selectron velocity,
the amplitude of the intermediate state $e^-_L e^-_L \to \tilde{e}^-_L \tilde{e}^-_L$ behaves like 
$\beta$, while for the other two cases it goes like $\beta^3$.
The dependence of $\sigma_{E1}$ on $\delta_{LL}$ is shown in Fig.~\ref{eedelta}. 
With SUSY masses not much larger than $\sim 200$ GeV the signal is of order 
${\cal O}$(10$^{-2}$) fb for $\delta_{LL} > {\cal O}(10^{-1})$. 
In addition the cross section is practically angle independent and thus 
insensitive to angular (or tranverse momentum) cuts.

The phenomenological points of the SUSY parameter space corresponding to gaugino 
masses $(M_1,\;M_2)=(80,\;160)$ GeV or  $(100,\;200)$ GeV and to 
slepton masses $m_L=100-200$ GeV and $\delta_{LL}>10^{-1}$
(which implies $\Delta m^2>10^{3}$ GeV$^{2}$) can give in the 
$e^- e^-$ mode a detectable LFV signal ($e^-e^- \to \ell^- e^-$)  
although at the level of ${\cal O}(1-25)$ events/yr with 
${L}_0=100$ fb$^{-1}$. Higher sensitivity to the SUSY 
parameter space could be obtained with larger $L_0$. 
It is interesting to note that this light sparticle spectrum, 
which is promising for collider discovery, is also preferred 
by the electroweak data fit. In Ref.~\cite{Altarelli} it is 
shown that light sneutrinos, charged left sleptons and light gauginos 
improve the agreement among the electroweak precision measurements 
and the lower bounds on the Higgs mass.
\EPSFIGURE[ht]
{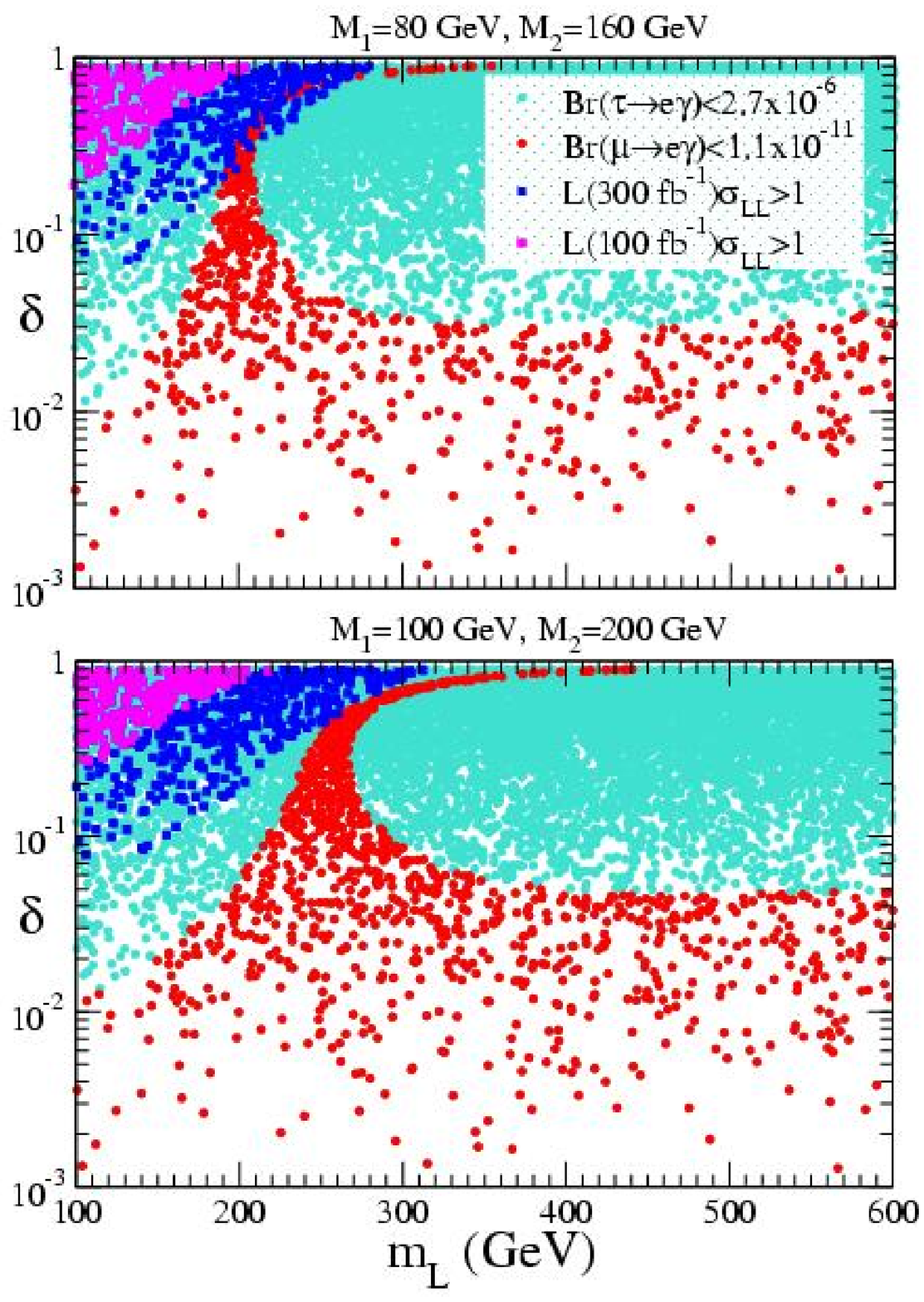, width=9cm,height=11cm,clip=}
{Scatter plot in the plane ($\delta_{LL}, m_L$) of: (a) the 
experimental bounds from $\mu\to e\gamma$ and 
$\tau \to \mu \gamma$ (allowed regions with circular dots); (b) regions where the signal 
can give at least one event with two different values of integrated 
luminosity (squared dots), for two sets of gaugino masses. Each signal point is calculated 
at $\sqrt{s}=2\tilde{m}_L$.\label{scaplot}}
On the other hand the experimental bounds on rare lepton 
decays $\mu ,\tau \to e\gamma$ set constraints on the LFV violating paramters $\Delta \tilde{m}^2$ or $\delta_{LL}$:
the upper bounds on the branching ratios define an allowed 
(and an excluded) region in the plane ($\delta_{LL}, m_L$) 
which are computed using the formulas given in Ref.~\cite{hisa2} (adapted to our model)  
for the LFV radiative lepton decays. These regions have to be compared 
with those satisfying the ``discovery'' condition
\be
L_0 \sigma(\delta_{LL}, m_L) \ge 1 .
\label{discoverycond}
\ee
Such a comparison is shown in Fig.~\ref{scaplot} from which it emerges
that: ($i$) 
For the $e^-e^- \to \ell^- e^-$ process 
there is an observable signal in the upper left corner of the  
($\delta_{LL}, m_L$) plane. The extension of this region depends
on $L_0$. 
($ii$) The bound from $\tau \to e \gamma$ does not 
constrain the region of the ($\delta_{LL}, m_L$) plane
compatible with an observable LFV signal and therefore
the reaction $e^-e^- \to \tau^- e^-$ could produce a detectable signal
whithin the highlighted regions of the parameter 
space (upper-left regions in the ($\delta_{LL}, m_L$) plane).
($iii$) As regards the constraints from the $\mu \to e \gamma$ decay
the allowed region in the ($\delta_{LL}, m_L$) plane is shown by 
the circular dark dots (red with colour): 
the process $e^-e^- \to \mu^- e^-$ is 
observable only in a small section of the parameter space
since the allowed region from the $\mu \to e \gamma$ decay 
almost does not overlap with the collider ``discovery'' region except 
for a very small fraction in the case of gaugino masses ($M_1=80$ GeV and 
$M_2=160 $ GeV). The compatibility of values of 
$\delta_{LL} \approx 1$ is due to a cancellation among the  
diagrams that describe the $\ell \to \ell' \gamma$ decay in particular points 
of the parameter space.

\section{Standard model background}    
These signals have the unique characteristic of a back to back high 
energy lepton pair and no missing energy. Sources of background 
were qualitatively discussed in 
Ref.~\cite{heusch}.
Here we discuss the reaction $e^{-}e^{-} \to \nu_{e}\nu_{e} {W^{-}}^{*} {{W}^{-}}^{*}$
followed by the decays ${W^{-}} {{W}^{-}} \to 
\ell^{-} {\bar{\nu}}_{\ell}{\ell^{-}}' {\bar{\nu}}_{{\ell}'}$,
with four neutrinos and a like sign-dilepton pair that can be of the same or
different flavour. 
This appears to be the most dangerous background, as it produces two 
leptons and missing energy, and therefore it is analyzed in more detail.

Figure~\ref{backg} shows the total 
cross section $e^{-}e^{-} \to \nu_{e}\nu_{e} {W^{-}}{{W}^{-}}$ calculated 
with the {\scshape{CompHEP}} package~\cite{comphep}, that allows to 
compute numerically the $17$ Feynman diagrams contributing at tree level.
Above the threshold for $W^{-}W^-$ gauge boson
production the cross section rises rapidly by orders of magnitude, 
becoming almost constant at high energies. 
In the region $\sqrt{s}\simeq 250-400$ GeV it increases from $10^{-2}$ 
fb to $1$ fb. In order to get an estimate of the cross section for   
the six particle final state process, the cross section 
$\sigma(e^{-}e^{-} \to {W^{-}}{{W}^{-}}\nu \nu)$ 
has to be multiplied by the branching ratio of the leptonic decays of 
the two gauge bosons, $\simeq 10\%$, so that  
$\sigma_{Background}\simeq 10^{-4}-10^{-2}$ fb, and it is at the level of the signal.
However the kinematical configuration of the final state 
leptons is completely different.    
Figure~\ref{backg} (upper-right) shows the angular distribution
of the gauge bosons which is peaked in the forward and backward 
directions so that the leptons produced by the $W$ gauge boson decay 
are emitted preferentially along the collsion axis.
\EPSFIGURE
{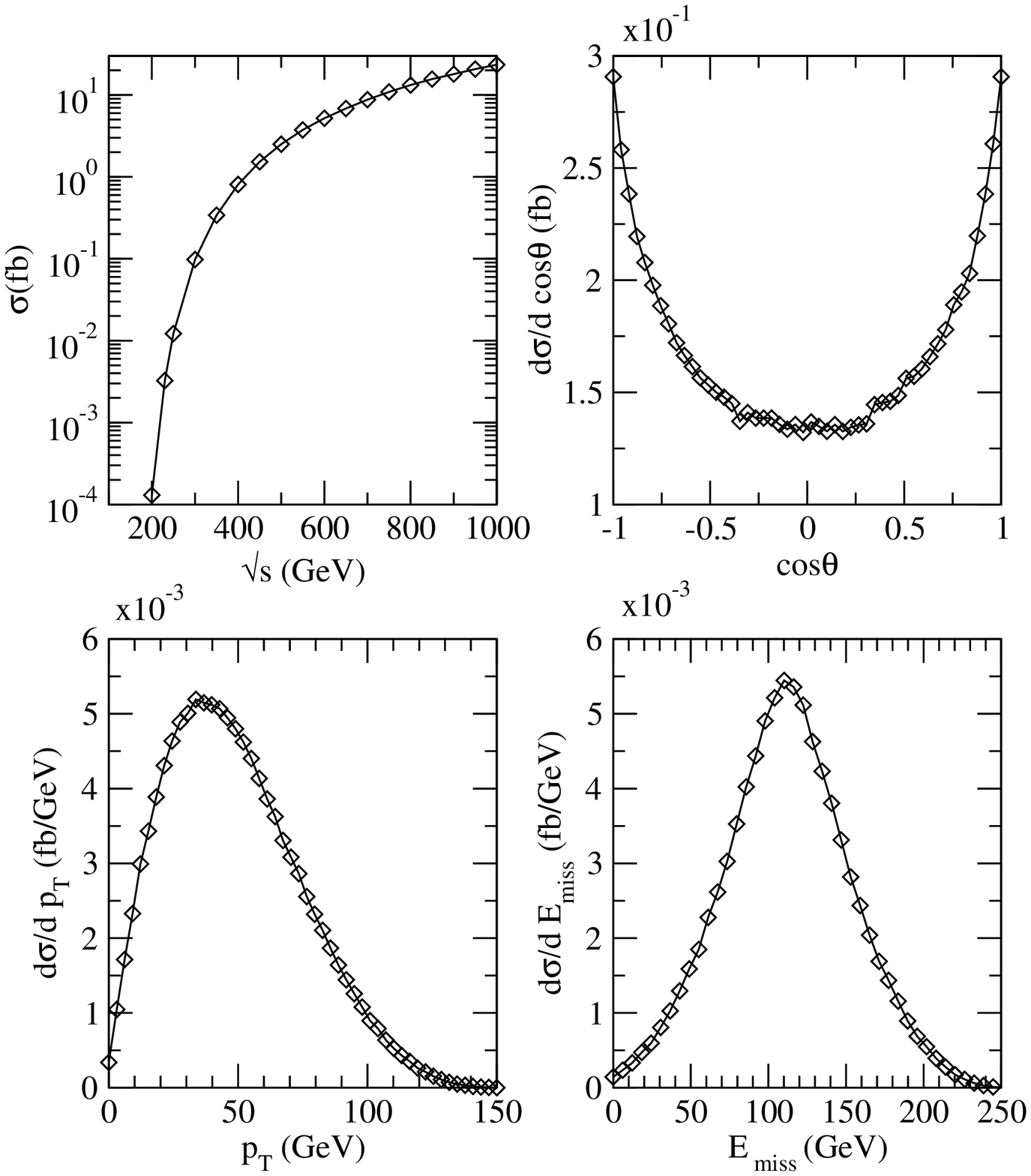, width=10cm,clip=}
{Total cross section and distributions for 
$e^{-}e^{-}\to W^{-}W^{-}\nu \nu$.
Upper-left figure: total cross section as a function of $\sqrt{s}$. 
Upper-right: angular distribution for 
a $W^{-}$ where $\theta$ is the angle among the collision axis and the boson momentum. 
Bottom-left: distribution of the transverse momentum of $W^{-}$.
Bottom-right: energy distribution of the two neutrinos. All distributions are calculated 
with $\sqrt{s}=300$ GeV.\label{backg}}
In addition their tansverse momenta will be softer compared to that of 
the signal: Fig.~\ref{backg} (bottom-left panel) shows that the 
transverse momenta distribution of the gauge bosons is peaked 
at $p^{P}_T=(\sqrt{s}/2-M_W)/2 \simeq 35$ GeV for 
$\sqrt{s}=300$ GeV. Consequentely the lepton distributions
will be peaked at $p^{P}_T/2 \simeq 17.5$ GeV.
The missing energy due to the undetected neutrinos
(Fig.~\ref{backg}, bottom-right panel) can be as large as 
$\simeq \sqrt{s} - 2M_W$. This distribution should be convoluted with
that of the neutrinos produced in the gauge boson decay.
Therefore it can be safely concluded that it will be possible to control 
this background because, with reasonable cuts on the transverse momenta of the leptons 
and on the missing energy, it will be drastically reduced, while - as mentioned
above - these cuts will not affect significantly the signal. 
The same conclusion holds for the signal with heavy Majorana neutrinos:
for $\sqrt{s} > 700$ GeV $\sigma(2\nu 2W)$ is 10 fb, the six particles final state has very
large missing energy carried away by neutrinos and charged leptons have soft distributions in transverse
momentum, thus cuts on missing energy and on the leptons $p_T$ reduce the background but not the signal.

\section{Conclusions}

The $e^- e^-$ option of the next generation of linear colliders offers 
the opportunity to test models of new physics through the discovery of lepton flavor violating 
signals, even if they arise only as a pure loop-level effect.
We have shown, using the maximum experimentally allowed mixings, that masses of heavy Majorana neutrinos 
up to $2-3$ TeV can be explored with the reaction $e^- e^- \to  \ell^- \ell^-, (\ell = \mu, \tau)$, because 
the amplitude gets an enhancement at the treshold for two gauge bosons production
and then shows a non-decoupling behaviour with the mass of the virtual heavy states.
For the similar reaction $e^- e^- \to  \ell^- e^-, (\ell = \mu, \tau)$ induced by slepton mixing in 
supersymmetric models,
in certain regions of the parameter space, the signal can reach the level of $10^{-2}$ fb around the 
threshold for selectrons pair production. 
The possibility of employing beams with high degree of 
longitudinal polarization is also essential to enhance the signal.  
On the other hand the standard model background is low and can be easily controlled.

\acknowledgments

M.~C.~thanks the organizers of AHEP 2003 Conference in Valencia for the financial support
which allowed his partecipation. The work of S.~K.~at INFN-Perugia in the period 2001-2002 has been supported by 
European Union under the Contract No.~HPMF-CT-2000-00752.

\end{document}